%% file: eBBH_method.tex
\pdfoutput=1
\documentclass[prd,
               twocolumn,
               amssymb,
               amsmath,
               eqsecnum,
               showpacs,
               letterpaper,
               superscriptaddress,
               altaffilletter,
               floatfix]
               {revtex4-1}

\usepackage{color}
\usepackage{graphicx}
\usepackage{latexsym}
\usepackage{float}
\usepackage{amsmath}
\usepackage{amssymb}
\usepackage{multirow}
\RequirePackage{fix-cm}

\begin{document}

\title{A Proposed Search for the Detection of Gravitational Waves from Eccentric Binary Black Holes}

\input{abstract}

\author{V.~Tiwari}
\affiliation{University of Florida, P.O.Box 118440, Gainesville, Florida, 32611, USA}
\author{S.~Klimenko}  
\affiliation{University of Florida, P.O.Box 118440, Gainesville, Florida, 32611, USA}
\author{N.~Christensen}
\affiliation {Carleton College, Northfield, MN 55057, USA}
\author{E.~A.~Huerta}
\affiliation{NCSA, University of Illinois at Urbana-Champaign, Illinois 61801, USA}
\author{S.~R.~P.~Mohapatra}
\affiliation {LIGO, Massachusetts Institute of Technology, Cambridge, MA 02139, USA}
\author{A.~Gopakumar}
\affiliation {Tata Institute for Fundamental Research, Mumbai 400005, India}
\author{M.~Haney}
\affiliation {Tata Institute for Fundamental Research, Mumbai 400005, India}
\author{P.~Ajith}
\affiliation {International Centre for Theoretical Sciences, Tata Institute of Fundamental Research, Bangalore 560012, India }
\author{S.~T.~McWilliams}
\affiliation {West Virginia University, Morgantown, WV 26506, USA }
\author{G.~Vedovato}  
\affiliation{INFN, Sezione di Padova, via Marzolo 8, 35131 Padova, Italy}
\author{M.~Drago} 
\affiliation{Max Planck Institut f\"{u}r Gravitationsphysik, Callinstrasse 38, 
             30167 Hannover, and Leibniz Universit\"{a}t Hannover, Hannover, Germany}
\author{F.~Salemi} 
\affiliation{Max Planck Institut f\"{u}r Gravitationsphysik, Callinstrasse 38, 
             30167 Hannover, and Leibniz Universit\"{a}t Hannover, Hannover, Germany}
\author{G.~A.~Prodi}    
\affiliation{University of Trento, Physics Department and INFN, Trento Institute 
for Fundamental Physics and Applications, via Sommarive 14, 38123 Povo, Trento, Italy}
\author{C.~Lazzaro}  
\affiliation{INFN, Sezione di Padova, via Marzolo 8, 35131 Padova, Italy}
\author{S.~Tiwari}  
\affiliation{University of Trento, Physics Department and INFN, Trento Institute 
for Fundamental Physics and Applications, via Sommarive 14, 38123 Povo, Trento, Italy}
\affiliation{Gran Sasso Science Institute (INFN), Via F. Crispi 7, I-67100, L’Aquila, Italy}
\author{G.~Mitselmakher}
\affiliation{University of Florida, P.O.Box 118440, Gainesville, Florida, 32611, USA}
\author{F.~Da Silva} 
\affiliation{University of Florida, P.O.Box 118440, Gainesville, Florida, 32611, USA}



\date[\relax]{Dated: \today }

\maketitle


\input{introduction}
\input{analysisoverview}
\input{results}
\input{discussion}

\input{acknowledgement}
\input{references}


\end{document}

%% file: abstract.tex
\begin{abstract}
Most of compact binary systems are expected to circularize before the frequency of emitted gravitational waves (GWs) enters the sensitivity band of the ground based interferometric detectors. 
However, several mechanisms have been proposed for the formation of binary systems, which retain eccentricity throughout their lifetimes. Since no matched-filtering algorithm has been developed 
to extract continuous GW signals from compact binaries on orbits with low to moderate values of eccentricity, and available algorithms to detect binaries on quasi-circular orbits are sub-optimal 
to recover these events, in this paper we propose a search method for detection of gravitational waves produced from the coalescences of eccentric binary black holes (eBBH). We study the search 
sensitivity and the false alarm rates on a segment of data from the second joint science run of LIGO and Virgo detectors, and discuss the implications of the eccentric binary search for the 
advanced GW detectors.
 
\end{abstract}

%% file: introduction.tex
\section{Introduction}

The existence of gravitational waves (GWs) is the direct consequence of linearized gravity. Several endeavors have been undertaken for the detection of GWs. Laster Interferometer Gravitational-wave 
Observatory (LIGO) and Virgo Observatory are ground based interferometric detectors built for this purpose \cite{LIGO}. Of the many sources these detectors are designed to detect, compact 
binary coalescence (CBC) are the most sought after emitters of gravitational waves \cite{LIGO}. Binaries formed with large orbital separations and at low frequencies are expected to circularize 
before they enter the sensitivity band of ground-based GW detectors. \cite{LIGOcbcRates}, however, several dynamical-formation scenarios support the formation and merger of binary systems while 
retaining eccentricity throughout their lifetime. For example, significant number of eccentric binary black holes (eBBH) can form within galactic nuclei through 2-body scattering. The presence 
of a super massive black hole (SMBH) can create steep density cusps of stellar mass black holes providing suitable environment for runaway encounters. If two BHs loose sufficient energy in such 
an encounter they will form a bound system and merge within hours of its formation \cite{lkl2009}. Another astrophysical scenario involves hierarchical triplets, modeled to consist of an inner 
and an outer binary. If the mutual inclination angle between the orbital planes of the inner and the outer binary is large enough, then the time averaged tidal force may induce oscillations in 
the eccentricity of the inner binary, known in the literature as the Kozai mechanism \cite{mh2002, fkr2000}. Some other formation scenarios have been proposed for the formation of eBBH, suggesting 
eBBH as a potential GW source for the ground-based detectors \cite{dlk2005, cb2005}. Overall, these scenarios suggest expected rate of coalescence detectable by advance LIGO to be 1-2500 per year. 
The eccentricity of these sources, when they become visible to the detectors, depends on their formation mechanism. An eBBH formed in a galactic core is expected to have eccentricity larger than 
0.9 at the time its orbital frequency is 5 Hz. On the contrary, eBBH formed in a globular cluster is expected to have low eccentricity and the eccentricity in the three body system is expected 
to oscillate. Detection of these sources offers rich information about their formation mechanics. Additionally, because of the high velocities involved during the periastron passage or presence 
of zoom-whirl behavior in the orbit can help probe into strong field regimes \cite{hl2009}. 

So far signals from merging stellar mass binary systems are searched mainly by matched-filtering the data with different families of templates. Separate searches have been conducted in the total 
mass range of 2 M$_\odot$ - 25 M$_\odot$ and 25 M$_\odot$ - 100 M$_\odot$ \cite{cbcS5low1, cbcS5low2, cbcS6low,cbcS6high1, cbcS6high2}. However, it has been shown that a large fraction of eBBH 
signals may be missed by the current template searches, which employ non-eccentric waveforms. Template searches are non-optimal for binaries with eccentricity more than 0.1 and currently only a 
dedicated search specifically targeting these systems can detect and study the rate of these sources \cite{mp1999, bz2010, hb2013}. 

Alternative methods have been proposed for the detection of eBBH (such as \cite{cm2015} for binaries with total mass less than 10 M$_\odot$). Moreover, there is an ongoing program that is developing 
a toolkit to detect and characterize eBBH events along with efforts in the GW modeling community \cite{hk2014, ya2009}. This paper is a significant step in that direction, in which, we describe an 
eBBH search using an excess energy (burst) method. These methods identify events by searching for a coincident appearance of the excess energy in two or more detector. Events surviving the consistency 
checks and criteria based on CBC model are admitted for further processing. Finally, we study the search sensitivity and the false alarm rates. We have performed a test search on the data obtained from 
the data obtained from the LIGO and Virgo detector from June of 2010 to October of 2010. Based on our results, we conclude, that the advanced detects can potentially detect eBBH signals, if the 
formation models predicting eBBH population hold true. In the event of null observation some of the models can be rejected at 90\% confidence.


The paper is organized as follows: Section \ref{detectors} contains information on the current ground based interferometric detectors. Section \ref{overview} presents an overview of the analysis. Section 
\ref{results} reports the results of the test search and we conclude with discussions in Section \ref{discuss}. 


%% file: analysisoverview.tex
\section{Detectors \label{detectors}}

The LIGO and Virgo detectors are kilometer scale, ground-based GW detectors. LIGO detectors are located at Livingston, LA (L1) and Hanford, WA (H1) and Virgo detector is located at Casina, Italy (V1). 
Before the sixth scientific run, there was also a second detector operating in Hanford (H2). So far the detectors have conducted two joint runs and the last run ended in 2010. Since then the detectors 
underwent a period of upgrade to increase their sensitivity. As of October 2015 LIGO Livingston and Hanford detectors have started collecting data with around four fold increase in the sensitivity. 
The Virgo detector is expected to begin operation within the next year. The design sensitivity of the advanced detectors, expected to be achieved by the year 2019, is approximately an order of magnitude 
larger \cite{advLIGO}.

We performed a test eBBH search on the segment of data collected by the LIGO and Virgo detectors from June of 2010 to October of 2010 (S6D). We used three-fold L1-H1-V1 network to perform the search. The 
detector's output is affected by variety of noise sources of the environmental and instrumental origin, hence, only a subset of the original data surviving the data quality flags was used in the search 
\cite{LIGOc2010, LIGOi2010}. Data quality flags are classified into different categories. Data generated during detector malfunction or when the coupling between detector output and noise source is 
well understood is flagged. Data surviving these flags is searched for GWs. Events crossing a pre-selected threshold are saved for processing. Further, event-by-event flags are applied on the saved 
events. Flags where coupling between detector noise and noise source is not well understood are applied. Another set of data quality flags is used to remove events with weak environmental and instrumental 
correlations \cite{s2011}.

\section{Analysis Overview \label{overview}}

\subsection{Data Analysis Algorithm}
The eBBH search uses the coherent waveburst (cWB) method \cite{kymm2008} to identify candidate events. cWB has been used for several burst searches on the S5-VSR1 and S6-VSR2/3 data \cite{bursts5, bursts6, lvc2012, lvc2014}. 
Recently cWB has been upgraded extensively~\cite{cwb2g}. Use of multi-resolution Wilson-Daubechies transforms \cite{nkm2012} was implemented to maximize collection of SNR from an event, resulting in recovery 
of more than 90\% of the signal-to-noise ratio (SNR) for binaries with total mass 20 M$_\odot$ or more. cWB performs multi-resolution time-frequency (TF) analysis of the detector data and searches for 
coincidental appearance of excess energy in the network data. Excess energy TF pixels are extracted and are clustered to form an event. Thereafter, a likelihood analysis is performed on the event and the 
expected signal is reconstructed. Two major coherent statistics are obtained during the likelihood analysis. Overall consistency of the event is quantified by the network correlation coefficient (cc). The GW events are 
expected to have a cc value close to unity, its maximum possible value. The strength of an event is quantified by the coherent network amplitude ($\eta$). $\eta$ is proportional to the SNR of the event.   

We employ two model based constraint to suppress background induced events. The eBBH signals are expected to have elliptical polarization. Events that do not satisfy this constraint are discarded \cite{cwb2g}. 
It is possible to identify significant events at this stage but as the eBBH signals are expected to have a chirping TF signature, only events with a chirping TF signature are admitted for further processing. 
The selection is made based on the reconstructed chirp mass and two goodness of fit parameters (``energy fraction'' and ``ellipticity'') of the events. These parameters are estimated during the processing of 
the data \cite{tk2015}. 

\subsection{Background Estimation}

\begin{figure}[htp]
  \centering
   \includegraphics[width=.45\textwidth, height=2.5in]{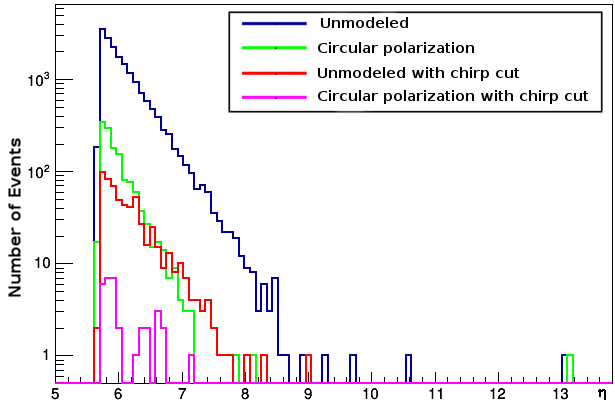}
   \caption{Distribution of coherent network amplitude, $\eta$, for the time shift-analysis performed over S6D (LHV) network.} 
   \label{eff-of-const}
\end{figure}

The false alarm rate of the background induced events are estimated by identifying events after performing relative time-slides between the data from different detectors. Data is shifted by a duration which is much 
longer than the maximum travel time of a GW between detectors, thereby removing existence of any coincident GWs. The cWB parameter space is not constrained, because of its “eyes wide open approach” 
and allows for the detection of multiple types of signals. Because of that, the cWB method is affected by background noises. Background induced events are suppressed by the application of elliptic polarization 
and reconstructed chirp mass constraint. Effect of these constraint is shown in Figure \ref{eff-of-const}. The number of background events are reduced by three orders of magnitude.

\section{Simulation}

The search sensitivity is quoted as the visible volume for the eBBH mergers. Simulation studies are performed by processing the data injected with the eBBH waveforms. In the eBBH search we introduce an astrophysical 
model to populate the parameter space from which injection parameter are randomly derived. This gives an opportunity to quantify the sensitivity of the search based on the parameters defining the astrophysical model. 
The range of these parameters can be constrained by using the results obtained from the eBBH search.

\subsection{Formation Model}

It is now well established that SMBHs are ubiquitous in the nuclei of nearby galaxies \cite{mmhgndi1995}. Theoretical studies suggest that mass segregation from individual objects and satellite stellar systems may 
lead to a high density of compact objects in the galactic center cusp \cite{bw1976, bw1977}. Multiple work show that black holes with mass around 10 M$_\odot$ should segregate to the inner parsec 
\cite{m1993, ha2006, mg2000, fak2006}.This dense stellar environment is ideally suited for the formation of eBBH, when runaway encounters can result is capture and quick merger of the binary. We model this 
astrophysical motivated scenario by implementing the model proposed by Kocsis and Loeb \cite{lkl2009}. 

The parameter defining the astrophysical model is the mass of the SMBH and $\beta$, which defines the mass function as
\begin{equation}
\frac{\text{d} N}{\text{d}m}\propto m^{-\beta}\,,
\label{mass_function}
\end{equation}
where $N$ is the number of binaries with mass $m$. The radial distribution of the stellar mass black 
holes in the galactic cusp is influenced the mass range of the stellar mass black holes. For the example eBBH search we fixed the mass range  from 5 M$_\odot$ to 25 M$_\odot$. Events are generated for the provided 
radial and mass distribution of the stellar mass black hole inside the cusp and mass of the SMBH (events are generated based on Equation 31 of \cite{lkl2009}, we do not consider other corrections considered in the 
paper). Once a binary is formed, periastron distance ($r_p$) is evolved in time, until the burst frequency, defined as, 
\begin{equation}f_{\text{avg}} = \frac{1}{\pi M}\frac{1}{\sqrt{2}r_p^{3/2}},\end{equation}
is achieved, where $M$ is the total mass of the binary.

\subsection{Simulated Waveforms}

We are using a waveform model that describes the inspiral, merger and ringdown of compact binaries on eccentric orbits \cite{emlp2013}. The model is based on mapping the binary to an effective single black hole system 
described by a Kerr metric, thereby including certain relativistic effects such as zoom-whirl-type behavior. Mass and total angular momentum of the binary are identified with the mass and spin parameters of the effective 
Kerr space time and the orbital angular momentum and energy with that of the geodesics. The parameters are evolved with dissipation coming from the quadrupole radiation term. This approach has the advantage of reproducing 
the correct orbital dynamics in the Newtonian limit and general-relativistic test particle limit, while still incorporating strong-field phenomena such as pericenter precession, frame dragging  and the existence of unstable 
orbits and related zoom-whirl dynamics. The waveforms provide a reasonably good agreement with results from numerical simulations \cite{bbcktm2008, kbbmc2011}.

\subsection{Visible Volume}

The visible volume, also called the sensitive volume, of the search is defined as,
\begin{equation} V_{\text{vis}}(m_1, m_2, e, r_p, \eta) = 4\pi \int_0^{\infty} \epsilon(m_1, m_2, e, r_p, \eta)r^2 \text{d}r \label{vis_vol},\end{equation}
where $\epsilon$ is the detection efficiency of the search. Using following equations,
\begin{equation} \epsilon = \text{d}N_{\text{det}}/\text{d}N_{\text{inj}}, \frac{1}{\rho_i} = \frac{4\pi r_i^2}{\text{d}N_{\text{inj}}/\text{d}r}, \end{equation}
where $N_{\text{inj}}$/$N_{\text{det}}$ is the number of injected/detected injections, $\text{d}N_{\text{inj}}/\text{d}r$ is the radial injection density and $\rho_i$ is the volume density, Equation \ref{vis_vol} 
becomes
\begin{equation} V_{\text{vis}} = \sum_i^{N_{\text{det}}} \frac{1}{\rho_i}.\label{vis_vol_formula}\end{equation}
The index $i$ runs over all the recovered injections.


\subsection{False Alarm Rate Density and Event Significance}

The significance of a foreground event can be determined by estimating its false alarm rate, defined as 
\begin{equation} \text{FAR}(\eta) = \frac{N(\eta)}{T}, \label{eq_far}\end{equation}
where $\eta$ is event's coherent network amplitude, $T$ is the accumulated livetime and $N(\eta)$ are the number of background events with coherent network amplitude greater than $\eta$. However, FAR values 
can not be used to compare significance of events across different networks. Searches can be combined by using the False Alarm Rate Density (FAD) statistic, which is defined as,
\begin{equation} \text{FAD}(\eta_j) = \text{min}\left( \frac{\text{FAR}(\eta_j)}{ V_{\text{vis}}(\eta_j)}, \frac{\text{FAR}(\eta_{j-1})}{ V_{\text{vis}}(\eta_{j-1})}\right). \label{fad_unb}\end{equation}
Events are ranked based on their FAD values with significant events having lower FAD rates.

To determine the event's significance, its FAD rate is compared to the time-volume product of the combined
search given by 
\begin{equation}
 \nu = \sum_k T_{\text{obs},k}V_{\text{vis}}(\text{FAD}),
\end{equation}
where the index $k$ runs over all the detector networks.
The mean of number of such events produced from background noise is
\begin{equation} \mu(\text{FAD}) = \text{FAD}\times \sum_k T_{\text{obs},k}V_{\text{vis}}(\text{FAD}).\end{equation}
Assuming FAD of the background events follows Poisson distribution, the false alarm probability (FAP) is given by 
\begin{equation} \text{FAP}(\eta) = 1 - \sum_{n=0}^{N-1}\frac{\mu^n}{n!}\exp(-\mu). \label{eq_FAP}\end{equation}



%% file: results.tex
\section{Results \label{results}}
\begin{figure}[htp]
\begin{center} 
\includegraphics[width=.45\textwidth]{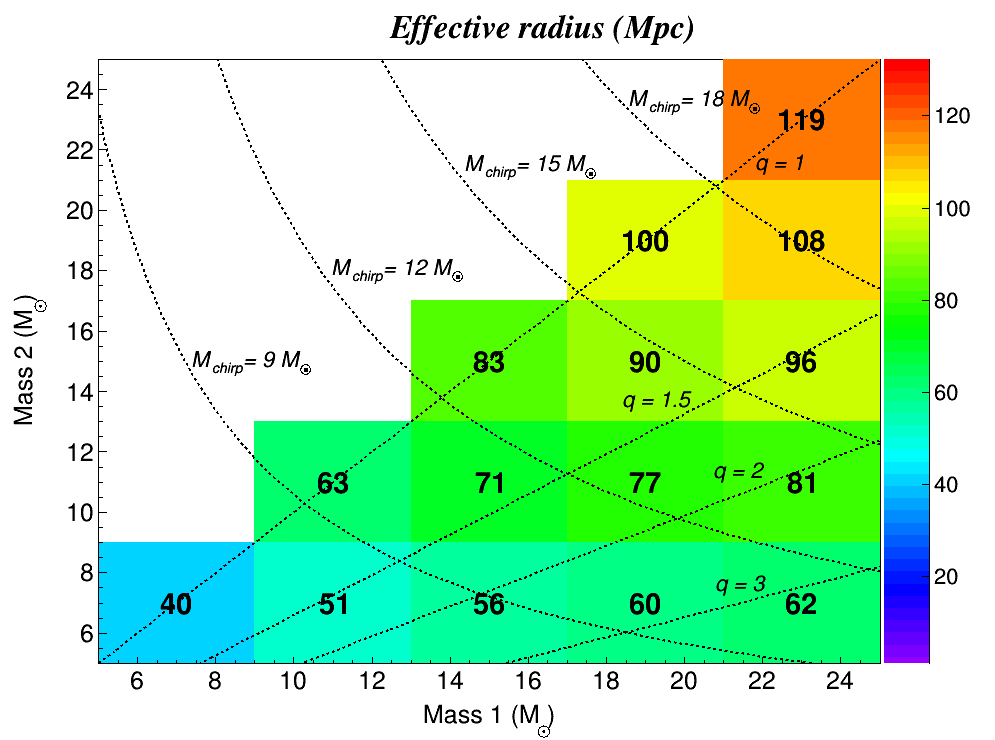} 
\caption{The effective range $R_{\text{eff}}$ in Mpc over component mass bin: S6D L1H1V1 network. The dotted contours represent constant mass ratio(q) and chirp mass($M_\text{chirp}$). Over 
all, the sensitive distance increases with the increase in the chirp mass and decrease in the mass ratio ($q$).}
\label{fig:effrad}
\end{center}
\end{figure}

In this section we discuss the projected sensitivity and expected rates for eBBH mergers based on the results obtained in the test run. Figure~\ref{fig:effrad} shows the sensitive distance for eBBH 
sources as a function of the component masses. Only recovered injections with FAR value of once in five years or less have been used to estimate the sensitive distance. The corresponding visible volume, estimated 
to be $\sim10^7$Mpc$^3$, is expected to increase by more than three orders of magnitude for advanced detectors. The eBBH merger rate of $\sim 10^{-9}$ per galactic nuclei, when averaged over SMBH density, 
results in eBBH coalescence rate of $\sim 10^{-10} $Mpc$^{-3}$ \cite{ar2002, t2013}. With these numbers, advanced detectors are expected to observe an average of one detection per observation year. There 
are astrophysical models projecting per galactic merger rate to be as high as $\sim 10^{-5}$ (including a factor of $\sim 30$ due to variance in the central number density of BHs). If these models hold 
true we expect to detect multiple eBBH signals with the advanced detectors. On the other hand, in the event of a null observation some of the astrophysical models can be rejected with confidence 
\cite{bcw2003, fb2007}. 

\begin{figure}[htp]
  \centering
   \includegraphics[width=.45\textwidth]{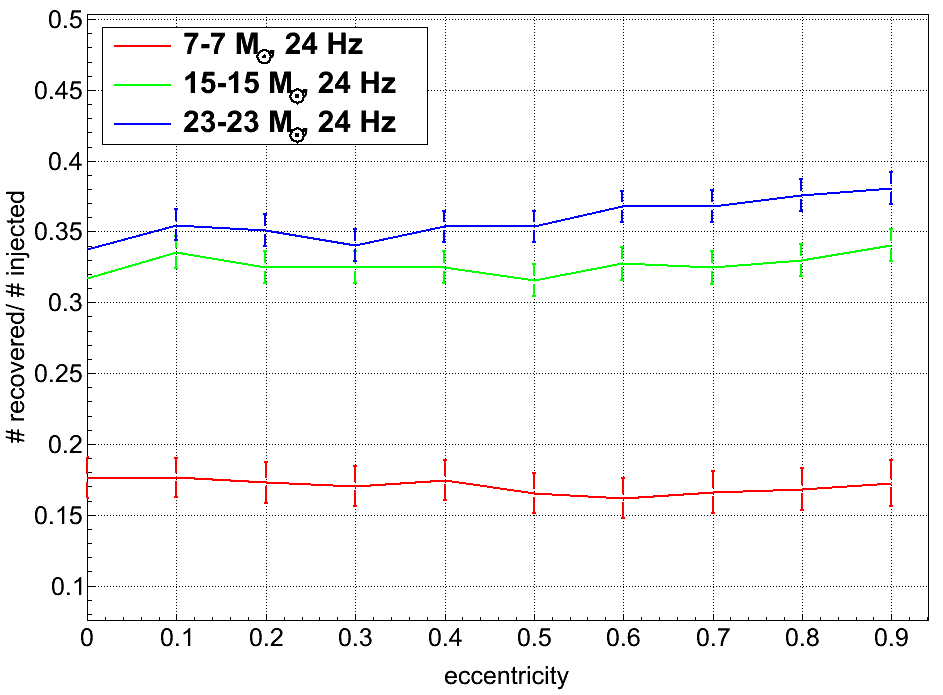}
   \caption{Efficiency vs eccentricity, for eccentricity values are at an orbital frequency of 48 Hz. Injection were made for three different masses. Heavier binaries show minor increase in efficiency with 
   increasing eccentricity (efficiency values depend on the chosen injection distance)). There is no visible trend for lower mass binaries.}  
   \label{fig:eff_ecc}
\end{figure}

A matched filtering search is an ideal choice for CBC sources, however, we can comment on the approximate fraction of eBBH signals proposed search 
can detect. Figure~\ref{fig:eff_ecc} plots the efficiency as a function of the eccentricity of binary at orbital frequency of 48 Hz. Efficiency is defined as the number of recovered injections divided 
by the number of injections. The injections have a fixed sky location. The efficiency does not show a visible trend for lower mass binaries. As expected, heavier binaries show minor increase in efficiency 
with increasing eccentricity (increased contribution from higher order modes). The search leaves the parameter space unconstrained in eccentricity, hence, the proposed eBBH search will also detect circular 
binaries with approximately equal efficiency. The effective radius for the example run is approximately 80\% of the matched filtering search \cite{cbcS6high1,cbcS6high2} performed for circular binaries. 
Hence, we expect the proposed search to recover half of the events, which could have been otherwise recovered by a matched filtering search using accurate waveforms of binaries on eccentric orbits. 

%% file: discussion.tex
\section{Discussion \label{discuss}}

We have introduced a novel search focused at the detection of GWs from eccentric binary black hole mergers. The search uses cWB algorithm to identify the events. A time-shift analysis is 
performed to estimate the background and simulation are performed to estimate sensitivity of the search. The search can use model based constraints, such as, polarization constraint and reconstructed 
chirp mass constraint to suppress the background. We show that these constraints suppress the background by three orders of magnitude. We describe FAD statistic which can be used to rank the events 
according to their significance.

We performed an example run and based on the obtained results we conclude that advanced detectors will detect multiple eBBH signals if the proposed astrophysical models hold true. The search will 
detect approximately half of the events a matched filter search would have detected. The search employs astrophysical model to populate the parameters space providing the opportunity to gauge the 
sensitivity of the search in terms of the parameters defining the astrophysical model. Hence, in the event of null observation it will become possible to reject some of the optimistic models.

%% file: acknowledgement.tex
\section{Acknowledgement}
We are thankful to the  National Science Foundation for support under grants PHY 1205512, PHY 1505308 and PHY 1505373. This document has been assigned LIGO Laboratory document number P1500171. We acknowledge support from the LIGO scientific 
and Virgo collaborations for providing the data used in the study.